\documentstyle[prd,aps,preprint,tighten,epsfig]{revtex}

\begin{document}

\draft

\title{Radiative Generation of $\theta_{13}$ with the Seesaw Threshold Effect}
\author{{\bf Jian-wei Mei} ~ and ~ {\bf Zhi-zhong Xing}}
\address{CCAST (World Laboratory), P.O. Box 8730, Beijing 100080, China \\
and Institute of High Energy Physics, Chinese Academy of Sciences, \\
P.O. Box 918 (4), Beijing 100039, China
\footnote{Mailing address} \\
({\it Electronic address: jwmei@mail.ihep.ac.cn; xingzz@mail.ihep.ac.cn}) }
\maketitle

\begin{abstract}
We examine whether an appreciable value of the lepton flavor mixing
angle $\theta_{13}$ at the electroweak scale $\Lambda_{\rm EW}$ can be 
radiatively generated from $\theta_{13} = 0^\circ$ at the GUT scale
$\Lambda_{\rm GUT}$. It is found that the renormalization-group running 
and seesaw threshold effects may lead to $\theta_{13} \sim 5^\circ$ at
low energies for two simple large-maximal mixing patterns of the MNS 
matrix in the minimal supersymmetric standard model. If $\theta_{12}$ 
is sufficiently large at $\Lambda_{\rm GUT}$, it will be possible to 
radiatively produce $\theta_{13} \sim 5^\circ$ at $\Lambda_{\rm EW}$
both in the standard model and in its supersymmetric extensions. The 
mass spectrum of three heavy right-handed Majorana neutrinos and the 
cosmological baryon number asymmetry via leptogenesis are also calculated.
\end{abstract}


\newpage

\section{Introduction}

The recent solar \cite{SNO}, atmospheric \cite{SK}, reactor 
(KamLAND \cite{KM} and CHOOZ \cite{CHOOZ}) and accelerator (K2K \cite{K2K})
neutrino oscillation experiments have provided us with very robust evidence 
that neutrinos are massive particles and their mixing involves two 
large angles ($\theta_{12} \sim 33^\circ$ and $\theta_{23} \sim 45^\circ$) 
and one small angle ($\theta_{13} < 13^\circ$). How small $\theta_{13}$
is remains an open question, but a global analysis of the presently 
available neutrino oscillation data \cite{FIT} indicates that 
$\theta_{13}$ is most 
likely to lie in the range $4^\circ \leq \theta_{13} \leq 6^\circ$. One
important target of the future neutrino experiments is just to measure 
$\theta_{13}$ \cite{T13}.

The smallness of $\theta_{13}$ requires a good theoretical reason, 
which might simultaneously account for the largeness of $\theta_{12}$ and 
$\theta_{23}$. If $\theta_{13} =0^\circ$ held, there should exist a kind 
of new flavor symmetry which forbids flavor mixing between the first
and third lepton families. While such a new symmetry is unlikely to 
exist at or below the electroweak scale ($\Lambda_{\rm EW} \sim 10^2$ GeV),
it might show up at a superhigh scale -- e.g., 
the scale of grand unified theories ($\Lambda_{\rm GUT} \sim 10^{16}$ GeV). 
Then a natural way to break this flavor symmetry and obtain 
$\theta_{13} \neq 0^\circ$ in a specific model is to run relevant 
parameters of the model from $\Lambda_{\rm GUT}$ to $\Lambda_{\rm EW}$ 
by making use of the renormalization group equations (RGEs) \cite{RGE} 
and taking account of the seesaw threshold effects \cite{Ratz}, either
in the standard model (SM) or in the minimal supersymmetric standard
model (MSSM). Antusch {\it et al} have
recently presented two simple examples (one with 
$\theta_{12} = \theta_{23} = 45^\circ$ and $\theta_{13} = 0^\circ$ at 
$\Lambda_{\rm GUT}$ \cite{T1}, and the other with 
$\theta_{12} = \theta_{13} = 0^\circ$ and 
$\theta_{23} = 45^\circ$ at $\Lambda_{\rm GUT}$ \cite{T2}) to radiatively 
generate $\theta_{12} \sim 33^\circ$ and $\theta_{13} \neq 0^\circ$, but 
their primary interest is in $\theta_{12}$ and their results for
$\theta_{13}$ are far below the best-fit values of 
$\theta_{13}$ obtained from the global analysis \cite{FIT}. Although the
low-scale output of $\theta_{13}$ is somehow adjustable by scanning the
parameter space of a given model at the GUT scale, we find that it is 
highly nontrivial to obtain $\theta_{13}(\Lambda_{\rm EW}) \sim 5^\circ$ 
from $\theta_{13}(\Lambda_{\rm GUT}) =0^\circ$ and fit all experimental 
data of neutrino oscillations in the meantime.

The main purpose of this paper is to examine whether an appreciable 
magnitude of $\theta_{13}$ can be radiatively generated, 
from $\Lambda_{\rm GUT}$ to $\Lambda_{\rm EW}$, through the seesaw 
thresholds. We shall consider four instructive patterns of the 
Maki-Nakagawa-Sakata (MNS) lepton mixing
matrix $V_{\rm MNS}$ at $\Lambda_{\rm GUT}$ as typical examples:
\begin{eqnarray}
{\rm Pattern ~ (A)}: ~~~~~~~~~~~~~~~~~~~ && V_{\rm MNS}
\; =\; \left ( \matrix{
~ 1	& ~ 0	& ~ 0 \cr
~ 0	& ~ \frac{\sqrt{2}}{2}	& ~ \frac{\sqrt{2}}{2} \cr
~ 0	& ~ - \frac{\sqrt{2}}{2}	& ~ \frac{\sqrt{2}}{2} \cr} 
\right ) P_\delta \; ,
\nonumber \\
{\rm Pattern ~ (B)}: ~~~~~~~~~~~~~~~~~~~ && V_{\rm MNS}
\; =\; \left ( \matrix{
\frac{\sqrt{2}}{2}	& ~ \frac{\sqrt{2}}{2}	& ~ 0 \cr
- \frac{1}{2}	& ~ \frac{1}{2}	& ~ \frac{\sqrt{2}}{2} \cr
\frac{1}{2}	& ~ - \frac{1}{2}	& ~ \frac{\sqrt{2}}{2} \cr} 
\right ) P_\delta \; ,
\nonumber \\
{\rm Pattern ~ (C)}: ~~~~~~~~~~~~~~~~~~~ && V_{\rm MNS}
\; =\; \left ( \matrix{
\frac{\sqrt 6}{3}	& \frac{\sqrt{3}}{3}	& 0 \cr
- \frac{\sqrt{6}}{6}	& \frac{\sqrt{3}}{3}	& \frac{\sqrt{2}}{2} \cr
\frac{\sqrt{6}}{6}	& - \frac{\sqrt{3}}{3}	& \frac{\sqrt{2}}{2} \cr}
\right ) P_\delta \; ,
\nonumber \\
{\rm Pattern ~ (D)}: ~~~~~~~~~~~~~~~~~~~ && V_{\rm MNS}
\; =\; \left ( \matrix{
\frac{\sqrt 3}{2}	& \frac{1}{2}	& 0 \cr
- \frac{\sqrt{2}}{4}	& \frac{\sqrt 6}{4}	& \frac{\sqrt{2}}{2} \cr
\frac{\sqrt{2}}{4}	& - \frac{\sqrt 6}{4}	& \frac{\sqrt{2}}{2} \cr}
\right ) P_\delta \; ,
\end{eqnarray}
where $P_\delta \equiv {\rm Diag} \{ 1, 1, e^{i\delta}\}$ with $\delta$
being a CP-violating phase in the standard parametrization of 
$V_{\rm MNS}$ \cite{FX01}
\footnote{For simplicity and illustration, we only take account of
a single CP-violating phase in this work.}.
While patterns (A) and (B) with CP conservation 
(i.e., $\delta =0^\circ$) 
have been discussed in Refs. \cite{T1} and \cite{T2}, we shall show 
that $\delta$ can actually play an important role in the radiative 
generation of $\theta_{13}$. Patterns (C) \cite{WOL} and 
(D) \cite{GPX} are phenomenologically favored to account for
current experimental data of solar and atmospheric neutrino oscillations. 
We find that the RGE running and seesaw threshold effects may allow us 
to obtain $\theta_{13} \sim 5^\circ$ at low energies from both
patterns (C) and (D) in the MSSM.
If $\theta_{12} \sim 60^\circ$ is taken at $\Lambda_{\rm GUT}$, it will 
be possible to radiatively produce $\theta_{13} \sim 5^\circ$ at
$\Lambda_{\rm EW}$ both in the SM and in the MSSM. As a by-product,
the mass spectrum of three heavy Majorana neutrinos and the
cosmological baryon number asymmetry via leptogenesis are also calculated.

\section{RGE running and threshold effects}

Let us make a simple modification of the SM by 
introducing three heavy right-handed neutrinos $N_i$ (for $i=1,2,3$)
and keeping the Lagrangian of electroweak interactions invariant 
under $\rm SU(2)_L \times U(1)_Y$ gauge transformation. In this
case, the Lagrangian relevant for lepton masses can be written as
\begin{equation}
-{\cal L}_{\rm lepton} \; =\; \bar{l}_{\rm L} Y_l e^{~}_{\rm R} H
+ \bar{l}_{\rm L} Y_\nu \nu^{~}_{\rm R} H^{\rm c} +
\frac{1}{2} \overline{\nu^{\rm c}_{\rm R}} M_{\rm R} \nu^{~}_{\rm R}
+ {\rm h.c.} \; ,
\end{equation}
where $l_{\rm L}$ denotes the left-handed lepton doublet; $e^{~}_{\rm R}$
and $\nu^{~}_{\rm R}$ stand respectively for the right-handed charged
lepton and Majorana neutrino singlets; and $H$ is the Higgs-boson
weak isodoublet (with $H^{\rm c} \equiv i\sigma^{~}_2 H^*$). If the
MSSM is taken into account,
one may similarly write out the Lagrangian relevant for lepton masses:
\begin{equation}
-{\cal L}'_{\rm lepton} \; =\; \bar{l}_{\rm L} Y_l e^{~}_{\rm R} H_1
+ \bar{l}_{\rm L} Y_\nu \nu^{~}_{\rm R} H_2 +
\frac{1}{2} \overline{\nu^{\rm c}_{\rm R}} M_{\rm R} \nu^{~}_{\rm R}
+ {\rm h.c.} \; ,
\end{equation}
where $H_1$ and $H_2$ (with hypercharges $\pm 1/2$) are the MSSM Higgs 
doublets. To obtain the effective (left-handed)
neutrino mass matrix, a common approach is to integrate $M_{\rm R}$ 
out of the full theory. This corresponds to a replacement of the
last two terms in ${\cal L}_{\rm lepton}$ or ${\cal L}'_{\rm lepton}$ 
by a dimension-5 operator, whose coupling matrix takes the well-known
seesaw form
$\kappa = -Y_\nu M^{-1}_{\rm R} Y^T_\nu$ \cite{SS}. However, the 
threshold effects in such a naive treatment have to be taken into 
account, because the mass eigenvalues of $N_i$ may have a strong 
hierarchy (for example, $M_3 \gg M_2 \gg M_1$). 

We take ${\cal L}_{\rm lepton}$ or ${\cal L}'_{\rm lepton}$
for granted at the GUT scale, where the Yukawa 
interactions of quarks and Higgs bosons with the coupling matrices
$Y_{\rm u}$ (up) and $Y_{\rm d}$ (down) can similarly be written out. 
To evolve the lepton mixing parameters from $\Lambda_{\rm GUT}$ to 
$\Lambda_{\rm EW}$ in a generic seesaw model 
\footnote{We assume the supersymmetry breaking scale $\Lambda_{\rm SUSY}$
to be close to the electroweak scale $\Lambda_{\rm EW}$,
just for the sake of simplicity. Even if 
$\Lambda_{\rm SUSY}/\Lambda_{\rm EW} \sim 10$ holds, the relevant RGE
running effects between these two scales are negligibly small for the
physics under consideration.},
one has to make use of a series of effective theories which are obtained 
by integrating out the heavy right-handed singlets $N_i$ step by step 
at their mass thresholds. The derivation of the one-loop RGEs and the
method for dealing with the effective theories have been presented
in Ref. \cite{Ratz} in an elegant way. Here we summarize a few essential 
steps to be taken in treating the seesaw threshold effects.

(a) We use the one-loop RGEs to run $Y_\nu$ and $M_{\rm R}$ from 
$\Lambda_{\rm GUT}$ to the heaviest right-handed neutrino mass scale 
$M_3$. A proper unitary transformation of the right-handed neutrino
fields allows us to diagonalize $M_{\rm R}$ at $M_3$ 
-- namely, $U^\dagger_{\rm R} M_{\rm R} U^*_{\rm R} = 
{\rm Diag}\{ M_1, M_2, M_3\}$. Then $Y_\nu$ is transformed into 
$Y_\nu U^*_{\rm R}$. The effective neutrino coupling matrix 
$\kappa^{~}_{(3)}$ can be obtained by integrating out $M_3$. 
In this case, we denote
\footnote{It should be noted that our notations (in particular,  
$\kappa = - Y_\nu M^{-1}_{\rm R} Y^T_\nu$) are somehow different from 
those of Ref. \cite{Ratz}, where the seesaw formula 
$\kappa = 2 Y^T_\nu M^{-1}_{\rm R} Y_\nu$ has been adopted.}  
\begin{equation}
Y_{\nu (3)} \; = \; Y_\nu U^*_{\rm R}
\left ( \matrix{
1 & 0 \cr
0 & 1 \cr
0 & 0 \cr} \right ) \; , ~~~~~~
\hat{Y}_{\nu (3)} \; = \; Y_\nu U^*_{\rm R}
\left ( \matrix{
0 \cr
0 \cr
1 \cr} \right ) \; , ~~~~~~
M_{\rm R (3)} \; = \; \left ( \matrix{
M_1 & 0 \cr
0 & M_2 \cr} \right ) \; 
\end{equation}
and get the tree-level matching relation
$\kappa^{~}_{(3)} = - \hat{Y}_{\nu (3)} M^{-1}_3 \hat{Y}^T_{\nu (3)}$,
where all variables have been set to the scale $\mu = M_3$. 

(b) We further run $Y_{\nu (3)}$, $M_{\rm R (3)}$ 
and $\kappa^{~}_{(3)}$ from $M_3$ down to the intermediate right-handed 
neutrino mass scale $M_2$. Because the RGE running effect may spoil the
diagonal feature of $M_{\rm R(3)}$, a re-diagonalization of 
$M_{\rm R (3)}$ at $M_2$ should be done by means of a $2\times 2$ unitary 
transformation matrix $\tilde{U}_{\rm R}$. Integrating out $M_2$, we 
arrive at
\begin{equation}
Y_{\nu (2)} \; = \; Y_{\nu (3)} \tilde{U}^*_{\rm R} 
\left ( \matrix{
1 \cr
0 \cr} \right ) \; , ~~~~~~~~~~
\hat{Y}_{\nu (2)} \; = \; Y_{\nu (3)} \tilde{U}^*_{\rm R} 
\left ( \matrix{
0 \cr
1 \cr} \right ) \; , ~~~~~~~~~~
M_{\rm R (2)} \; = \; \left ( \matrix{
M_1 \cr} \right ) \; ~
\end{equation}
and the tree-level matching condition
$\kappa^{~}_{(2)} = \kappa^{~}_{(3)}
- \hat{Y}_{\nu (2)} M^{-1}_2 \hat{Y}^T_{\nu (2)}$,
where all variables have been set to the scale $\mu = M_2$. 

(c) We follow a similar way to run $Y_{\nu (2)}$, 
$M_{\rm R (2)}$ and $\kappa^{~}_{(2)}$ from $M_2$ down to the lightest 
right-handed neutrino mass scale $M_1$. As $M_{\rm R (2)}$ is 
actually a $1\times 1$ mass matrix, it does not need to be 
re-diagonalized at $M_1$. Integrating out $M_1$, we obtain
\begin{equation}
\kappa \; \equiv \; \kappa^{~}_{(1)} \; = \; \kappa^{~}_{(2)}
- \hat{Y}_{\nu (1)} M^{-1}_1 \hat{Y}^T_{\nu (1)} \; ,
\end{equation}
where $\hat{Y}_{\nu (1)} = Y_{\nu (2)}$ holds, and all variables have 
been set to the scale $\mu = M_1$. 

(d) Finally, we run $\kappa$ 
from $M_1$ down to the electroweak scale $\Lambda_{\rm EW}$. The
one-loop RGE governing the evolution of $\kappa$ is given by \cite{RGE} 
\begin{equation}
16\pi^2 \frac{{\rm d} \kappa}{{\rm d} t} \; = \;
\alpha \kappa + C \left [ \left (Y_l Y^\dagger_l \right ) \kappa
+ \kappa \left (Y_l Y^\dagger_l \right )^T \right ] \; ,
\end{equation}
where $t = \ln (\mu/M_1)$ with $\mu$ being the renormalization
scale. We have $C = -1.5$, $\alpha \approx
-3g^2_2 + 6 f^2_t + \lambda$ in the SM and $C = 1$, 
$\alpha \approx -1.2 g^2_1 - 6g^2_2 + 6 f^2_t$ in the MSSM \cite{MX}, 
where $g^{~}_{1,2}$ denote the gauge couplings, $f_t$ denotes
the top-quark Yukawa coupling, and $\lambda$ denotes the
Higgs self-coupling in the SM. After spontaneous gauge 
symmetry breaking, we arrive at the fermion mass matrices
$M_\nu = v^2 \kappa$, $M_l = v Y_l$, 
$M_{\rm u} = v Y_{\rm u}$ and $M_{\rm d} = v Y_{\rm d}$ in the
SM; and $M_\nu = v^2 \kappa \sin^2\beta$, $M_l = v Y_l \cos\beta$, 
$M_{\rm u} = v Y_{\rm u} \sin\beta$ and 
$M_{\rm d} = v Y_{\rm d} \cos\beta$ in the MSSM, where 
$v \approx 174$ GeV stands for the vacuum expectation value of the
neutral Higgs field in the SM, and $\tan\beta$ represents the
ratio of two vacuum expectation values in the MSSM. 

\section{Initial conditions and assumptions}

The lepton (or quark) flavor mixing matrix 
$V_{\rm MNS}$ (or $V_{\rm CKM}$) arises from the 
mismatch between the diagonalizations of $Y_l$ (or $Y_{\rm u}$) and 
$\kappa$ (or $Y_{\rm d}$). Without loss of generality, we arrange
$Y_l$ and $Y_{\rm u}$ to be diagonal, real and positive at 
$\Lambda_{\rm GUT}$; i.e.,
\begin{equation}
Y_{\rm u} \; =\; \frac{1}{\Omega_1} \left ( \matrix{
m_u & 0 & 0 \cr
0 & m_c & 0 \cr
0 & 0 & m_t \cr} \right ) \; , ~~~~~~~~~~~~~~~~
Y_l \; =\; \frac{1}{\Omega_2} \left ( \matrix{
m_e & 0 & 0 \cr
0 & m_\mu & 0 \cr
0 & 0 & m_\tau \cr} \right ) \; , 
\end{equation}
where $\Omega_1 = \Omega_2 = v$ in the SM, and
$\Omega_1 = v\sin\beta$ and $\Omega_2 = v\cos\beta$ in the MSSM.
In this flavor basis, $Y_{\rm d}$ and $Y_\nu$ can generally be
expressed as 
\begin{equation}
Y_{\rm d} \; =\; \frac{1}{\Omega_2} V_{\rm CKM} \left ( \matrix{
m_d & 0 & 0 \cr
0 & m_s & 0 \cr
0 & 0 & m^{~}_b \cr} \right ) U_{\rm d} \; , ~~~~~
Y_\nu \; =\; y_\nu V_\nu \left ( \matrix{
r_1 & 0 & 0 \cr
0 & r_2 & 0 \cr
0 & 0 & 1 \cr} \right ) U_\nu \; , 
\end{equation}
where $r_1$, $r_2$ and $y_\nu$ are three real and positive 
dimensionless parameters characterizing the eigenvalues of $Y_\nu$; 
and $V_\nu$, $U_\nu$ and $U_{\rm d}$ are three $3\times 3$ unitary 
matrices. It is obvious that the complex phases of $V_\nu$ and $U_\nu$
can be re-arranged, such that the former contains a single irremovable 
CP-violating phase as $V_{\rm CKM}$ does. Furthermore, one may
redefine the relevant right-handed fields of quarks and neutrinos to 
rotate away $U_{\rm d}$ from $Y_{\rm d}$ and $U_\nu$ from $Y_\nu$.
Such a transformation of $Y_\nu$ is equivalent to absorbing $U_\nu$ 
into the Majorana mass matrix $M_{\rm R}$, which is not required to be
diagonal at $\Lambda_{\rm GUT}$.
Once $U_{\rm d}$ and $U_\nu$ are rejected, we are only left with seven 
unknown parameters in the above Yukawa coupling matrices (namely, three 
eigenvalues of $Y_\nu$ and four mixing parameters of $V_\nu$).

To fix the pattern of $M_{\rm R}$ at $\Lambda_{\rm GUT}$, we 
extrapolate the effective neutrino coupling matrix $\kappa$ and
the lepton flavor mixing matrix $V_{\rm MNS}$ up to the GUT scale:
\begin{equation}
\kappa \; =\; \frac{1}{\Omega^2_1} V_{\rm MNS} 
\left ( \matrix{
m_1 & 0 & 0 \cr
0 & m_2 & 0 \cr
0 & 0 & m_3 \cr} \right ) V^T_{\rm MNS} \; ,
\end{equation}
where $m_i$ (for $i=1,2,3$) denote the physical masses of three 
light neutrinos. Then $M_{\rm R}$ can be determined from the inverted 
seesaw formula $M_{\rm R} = -Y^T_\nu \kappa^{-1} Y_\nu$ \cite{XZ}.
Note that $V_{\rm MNS}$ consists of three mixing angles and three
CP-violating phases. For the sake of simplicity, here we only take 
account of the ``Dirac-like'' phase of $V_{\rm MNS}$. Then
the unitary matrices $V_{\rm CKM}$, $V_{\rm MNS}$ and $V_\nu$ may
universally be parametrized as
\begin{equation}
V \; =\; \left ( \matrix{
1 & 0 & 0 \cr
0 & c_{23} & s_{23} \cr
0 & -s_{23} & c_{23} \cr} \right ) 
\left ( \matrix{
c_{13} & 0 & s_{13} \cr
0 & e^{-i\delta} & 0 \cr
-s_{13} & 0 & c_{13} \cr} \right )
\left ( \matrix{
c_{12} & s_{12} & 0 \cr
-s_{12} & c_{12} & 0 \cr
0 & 0 & 1 \cr} \right ) \; ,
\end{equation}
where $c_{ij} \equiv \cos\theta_{ij}$ and $s_{ij} \equiv \theta_{ij}$.
In the limit of $\theta_{13} = 0^\circ$, the complex phase of 
$V_{\rm MNS}$ actually serves as a ``Majorana-like'' phase and may 
significantly affect the RGE running behaviors of neutrino masses and 
lepton flavor mixing angles \cite{Haba}. Four typical patterns of
$V_{\rm MNS}$, as already listed in Eq. (1), will be taken in our 
following investigation. 

To be specific, we only pay attention to the normal mass hierarchy of 
three light neutrinos (i.e., $m_3 > m_2 > m_1$). Then we arrive at
$m_2 = \sqrt{m^2_1 + \Delta m^2_{\rm sun}}~$ and
$m_3 = \sqrt{m^2_2 + \Delta m^2_{\rm atm}}~$, where $\Delta m^2_{\rm sun}$
and $\Delta m^2_{\rm atm}$ denote the mass-squared differences
of solar and atmospheric neutrino oscillations. Since the low-scale
values of $(m_u, m_c, m_t)$, $(m_d, m_s, m_b)$, $(m_e, m_\mu, m_\tau)$ 
and $(\Delta m^2_{\rm sun}, \Delta m^2_{\rm atm})$ are all known, 
we just have a single unknown mass parameter ($m_1$). On the other hand,
four parameters of $V_{\rm CKM}$ are also known at low 
energies \cite{PDG}. Given
a special pattern of $V_{\rm MNS}$ in Eq. (1), only its CP-violating
phase is not fixed. In short, we are totally left with nine free 
parameters at $\Lambda_{\rm GUT}$: the lightest neutrino mass $m_1$, 
three eigenvalues of $Y_\nu$, four mixing parameters of $V_\nu$ and 
the CP-violating phase of $V_{\rm MNS}$. We should also specify
the value of the Higgs mass $m^{~}_H$ in the SM and that of 
$\tan\beta$ in the MSSM when numerically solving the relevant RGEs.

Although $y_\nu$, $r_1$ and $r_2$ are arbitrary parameters, they are
expected to be of or below ${\cal O}(1)$. The unknown rotation and
phase angles of $V_\nu$ and $V_{\rm MNS}$ are allowed to take possible 
values in the range between $0$ and $2\pi$. As for those parameters 
whose sizes are known at low energies, one may properly adjust their 
initial values at $\Lambda_{\rm GUT}$ and run the RGEs to reproduce 
their low-scale values within reasonable error bars. We are then able
to fix the parameter space by fitting all relevant low-scale data. 
Such a phenomenological approach will allow us to examine whether 
$\theta_{13}(\Lambda_{\rm EW}) \sim 5^\circ$ can be generated 
from $\theta_{13}(\Lambda_{\rm GUT}) = 0^\circ$ for $V_{\rm MNS}$.
Naively, we speculate that an appreciable RGE enhancement of $\theta_{13}$ 
may take place either between the scales $\Lambda_{\rm GUT}$ and $M_1$ 
or between the scales $M_1$ and $\Lambda_{\rm EW}$. In either case,
the masses of three light neutrinos are required to be nearly degenerate.
As the seesaw threshold effect can significantly affect the RGE running
behaviors in most cases \cite{Ratz,T1,T2,King}, it is more likely 
that the dominant RGE enhancement of $\theta_{13}$ occurs between
$\Lambda_{\rm GUT}$ and $M_1$. We shall demonstrate this observation in 
our subsequent numerical calculations.

\section{Numerical examples and discussions}

We carry out a numerical analysis of the RGE running and seesaw 
threshold effects by following the strategies outlined above and
taking four typical patterns of $V_{\rm MNS}$ at the GUT scale. Our 
results are summarized in Table 1 and Figs. 1 and 2. Some comments
and discussions are in order.

(1) We have carefully examined the sensitivity of 
$V_{\rm MNS} (\Lambda_{\rm EW})$ to the value of every free parameter
at $\Lambda_{\rm GUT}$. We find that the inputs of $m_1$, 
$y_\nu$, $\theta^{\rm MNS}_{12}$, $\delta_{\rm MNS}$, 
$\theta_{13}$ of $V_\nu$ and $\Delta m^2_{\rm sun}$ appear to be very 
important in adjusting the output of $\theta^{\rm MNS}_{13}$  
at $\Lambda_{\rm EW}$, while the other
parameters mainly play a role in fine-tuning the results. In particular,
the effects of $r_1$, $r_2$, $\theta_{12}$ of $V_\nu$, $\theta_{23}$ of 
$V_\nu$ and $\delta$ of $V_\nu$ are insignificant, because the
dominant RGE enhancement of $\theta^{\rm MNS}_{13}$ 
takes place above the heaviest right-handed neutrino mass scale $M_3$. 
Hence we have simply fixed $\theta_{12} =\theta_{23} =\delta =0^\circ$ 
for $V_\nu$ at $\Lambda_{\rm GUT}$. Note that the initial values of 
$\delta_{\rm MNS}$ and $\theta_{13}$ of $V_\nu$ are important in 
specifying the running tendency of two neutrino mass-squared differences
and three mixing angles of $V_{\rm MNS}$, and they have to be
sufficiently large (as shown in Table 1) in order to generate an
appreciable magnitude of $\theta^{\rm MNS}_{13}$ at low energies. 
To be more explicit, the evolution of $\theta^{\rm MNS}_{12}$ strongly
relies on $\delta_{\rm MNS}$ in the fitting, while the 
enhancement of $\theta^{\rm MNS}_{13}$ (from zero at $\Lambda_{\rm GUT}$
to a few degrees at $\Lambda_{\rm EW}$) requires a big input value for 
$\theta_{13}$ of $V_\nu$ ($\sim 45^\circ$, for example). 
It is also worth remarking that the parameter space obtained here
should by no means be unique; or rather, it mainly serves for
illustration. To exhaustively explore the allowed ranges of
all relevant parameters is a quite lengthy work and will be 
presented elsewhere \cite{Mei}.

(2) Restricting ourselves to the typical parameter space illustrated 
in Table 1, we find that it is difficult to produce 
$\theta^{\rm MNS}_{13} \sim 5^\circ$ at $\Lambda_{\rm EW}$ in 
the SM. Our results yield $\theta^{\rm MNS}_{13} \sim 3^\circ$ for
pattern (A) and $\theta^{\rm MNS}_{13} \sim 1.5^\circ$ for 
patterns (B), (C) and (D) at low energies. 
In the MSSM, however, a more strong RGE enhancement of 
$\theta^{\rm MNS}_{13}$ becomes possible. For example, we obtain 
$\theta^{\rm MNS}_{13} \sim 3^\circ$ for pattern (B) 
with $\tan\beta \sim 10$ and $\theta^{\rm MNS}_{13} \sim 5^\circ$ 
for patterns (C) and (D) with $\tan\beta \sim 20$ at the electroweak 
scale. We think that the numerical results in the
supersymmetric case are encouraging for model building, because 
a new kind of flavor symmetry at $\Lambda_{\rm GUT}$ might naturally
lead to the bi-maximal mixing pattern (B) or the large-maximal mixing
patterns (C) and (D). 

(3) The running behavior of $\theta^{\rm MNS}_{12}$ deserves some
more remarks. As already noticed in Refs. \cite{RGE,Ratz,T1,T2}, 
the evolution of $\theta^{\rm MNS}_{12}$ from $\Lambda_{\rm EW}$
to $\Lambda_{\rm GUT}$ (or vice versa) depends sensitively on how big 
its initial value is and whether three light neutrinos have a near 
mass degeneracy. Figs. 1 and 2 illustrate that 
$\theta^{\rm MNS}_{12}(\Lambda_{\rm EW}) \sim 33^\circ$ can 
radiatively be obtained, either from 
$\theta^{\rm MNS}_{12}(\Lambda_{\rm GUT}) = 0^\circ$ in pattern (A)
or from $\theta^{\rm MNS}_{12}(\Lambda_{\rm GUT}) 
\sim (30^\circ - 45^\circ)$ in patterns (B), (C) and (D). 
For pattern (A) or (B), the RGE running of $\theta^{\rm MNS}_{12}$ 
in the MSSM is quite similar to that in the SM. The reason for this
similarity is simply that the magnitudes of $m_1$ (SM) and $m_1$ (MSSM) 
at $\Lambda_{\rm EW}$ are comparable ($\sim 0.05$ eV) and the value 
of $\tan\beta$ in the MSSM case is mild ($\sim 10$). When patterns
(C) and (D) are concerned, however, the running behavior of 
$\theta^{\rm MNS}_{12}$ in the MSSM is more violent than that in the
SM. The reason for such a remarkable difference is two-fold: first,
$m_1(\Lambda_{\rm EW}) \sim 0.14$ eV and
$m_1(\Lambda_{\rm GUT}) \sim 0.20$ eV in the MSSM case imply that 
the masses of three light neutrinos are nearly degenerate in the
entire running course -- this near mass degeneracy can significantly
affect the behavior of $\theta^{\rm MNS}_{12}$ \cite{RGE};
second, the large value of $\tan\beta$ ($\sim 20$) plays an important 
role in enhancing the RGE evolution of light neutrino masses and flavor 
mixing angles (e.g., 
$\dot{\theta}^{\rm MNS}_{12} \propto (1 + \tan^2\beta)$ \cite{RGE}).

(4) We stress that $\theta^{\rm MNS}_{13} \sim 5^\circ$ at 
$\Lambda_{\rm EW}$ can be radiatively generated from 
$\theta^{\rm MNS}_{13} = 0^\circ$ at $\Lambda_{\rm GUT}$ even
in the SM, only if we go beyond the four simple patterns considered 
above. A key point is to assume $\theta^{\rm MNS}_{12}$ to be 
large enough at the GUT scale. Such an example is presented in Fig. 3,
where $\theta^{\rm MNS}_{12} = 67^\circ$ (SM) versus
$\theta^{\rm MNS}_{12} = 63^\circ$ (MSSM) has been taken. One may
see that the dominant RGE suppression of $\theta^{\rm MNS}_{12}$
occurs from $\Lambda_{\rm GUT}$ to $M_3$, and the dominant RGE 
enhancement of $\theta^{\rm MNS}_{13}$ takes place in the same region.
Why $\theta^{\rm MNS}_{12} > 45^\circ$ holds at $\Lambda_{\rm GUT}$  
is certainly a big question. Putting aside this question, we remark
that $\theta^{\rm MNS}_{12} \sim 33^\circ$ at low energies can in
principle be produced from an arbitrary value of $\theta^{\rm MNS}_{12}$
at the GUT scale via the seesaw threshold effects. Furthermore, the
radiative generation of $\theta^{\rm MNS}_{13}$ is highly sensitive to 
the initial condition of $\theta^{\rm MNS}_{12}$. We observe that
the running of $\theta^{\rm MNS}_{23}$ is rather stable, unlike 
$\theta^{\rm MNS}_{12}$ and $\theta^{\rm MNS}_{13}$. Note that
the CP-violating phase $\delta_{\rm MNS}$ is also stable against
radiative corrections from $\Lambda_{\rm GUT}$ to $\Lambda_{\rm EW}$.
This conclusion is true for both the example under consideration and
the four patterns discussed above.

(5) A by-product of our analysis is the determination of three heavy
right-handed neutrino masses, as shown in Table 1. We see that 
they have a clear normal hierarchy. It is then possible to calculate
the cosmological baryon number asymmetry via leptogenesis \cite{FY}. 
Instead of describing the technical details of leptogenesis \cite{L}, 
we only give a brief summary of its essential points in the following.
Lepton number violation induced by the third term of
${\cal L}_{\rm lepton}$ in Eq. (1) or of ${\cal L}'_{\rm lepton}$ 
in Eq. (2) allows decays of three heavy Majorana neutrinos $N_i$
to happen: $N_i \rightarrow l + h$ and
$N_i \rightarrow \bar{l} + h^{\rm c}$, where $h = H$ in the SM or
$h = H^{\rm c}_2$ in the MSSM. Because each decay mode occurs at
both tree and one-loop levels (via the self-energy and vertex
corrections), the interference between these two decay amplitudes may
result in a CP-violating asymmetry $\varepsilon_i$ between
$N_i \rightarrow l + h$ and its CP-conjugated process.
If the interactions of $N_1$ are in thermal equilibrium when 
$N_3$ and $N_2$ decay, the asymmetries $\varepsilon_3$ and 
$\varepsilon_2$ can be erased before $N_1$ decays. Then only the 
asymmetry $\varepsilon_1$ produced by the out-of-equilibrium decay
of $N_1$ survives. This CP-violating asymmetry may lead to
a net lepton number asymmetry $Y_{\rm L} \propto \varepsilon_1$, 
and the latter is eventually converted into a net baryon number 
asymmetry $Y_{\rm B}$ via the nonperturbative sphaleron 
processes \cite{Kuzmin}: $Y_{\rm B} \approx -0.55 Y_{\rm L}$ in the
SM or $Y_{\rm B} \approx -0.53 Y_{\rm L}$ in the MSSM. We follow
these steps to evaluate $Y_{\rm B}$ for four patterns listed in Table 1. 
It turns out that only pattern (B) can yield an appreciable result,
which lies in the generous range
$0.7 \times 10^{-10} \lesssim Y_{\rm B} \lesssim 1.0 \times 10^{-10}$
drawn from the recent WMAP observational data \cite{WMAP}. Compared 
with pattern (B), patterns (A), (C) and (D) have relatively small 
values of $M_1$ ($\sim 10^9$ GeV) and $M_1/M_2$ ($\sim 3\times 10^{-2}$).
The outputs of $Y_{\rm B}$ in these three patterns are therefore
suppressed and below the observational result.
If the example shown in Fig. 3 is taken into account, we may obtain
$Y_{\rm B} \approx 7.7 \times 10^{-11}$ (SM) or
$Y_{\rm B} \approx 8.0 \times 10^{-11}$ (MSSM), which is compatible
with the WMAP data.

Finally, it is worthwhile to point out that we have also taken a 
look at the ``democratic'' neutrino mixing pattern \cite{FX}
\begin{equation}
V_{\rm MNS} \; =\; \left ( \matrix{
\frac{\sqrt{2}}{2} & \frac{\sqrt{2}}{2} & 0 \cr
-\frac{\sqrt{6}}{6} & \frac{\sqrt{6}}{6} & \frac{\sqrt{6}}{3} \cr
\frac{\sqrt{3}}{3} & -\frac{\sqrt{3}}{3} & \frac{\sqrt{3}}{3} \cr}
\right ) \;
\end{equation}
at the GUT scale and examined the possibility to radiatively produce
$\theta_{13} \sim 5^\circ$ at the electroweak scale and to 
simultaneously fit all experimental data of neutrino oscillations. 
We find that it is extremely difficult, if not impossible, to obtain
$\theta_{23} \sim 45^\circ$ at low energies. The reason is simply
that the initial value of $\theta_{23}$ ($\approx 54.7^\circ$) 
is not close to $45^\circ$ and the RGE evolution cannot significantly 
change the magnitude of this mixing angle. Therefore, we argue that
radiative corrections might not be a very natural way to break the 
lepton flavor democracy. The latter could be explicitly broken in some 
other possible ways at or below the GUT scale.

\section{Summary}

We have conjectured that the lepton flavor mixing angle $\theta_{13}$
might be vanishing at the GUT scale due to the existence of a kind of
new flavor symmetry. Starting from this point of view, we have
examined whether an appreciable value of $\theta_{13}$ at low energies
can be radiatively generated via the RGE running and seesaw threshold
effects. Four simple but typical patterns of the lepton flavor mixing
matrices have been considered as the initial conditions at 
$\Lambda_{\rm GUT}$. It is found that the dominant 
RGE enhancement of $\theta_{13}$ takes place from $\Lambda_{\rm GUT}$ 
to the heaviest right-handed neutrino mass scale. For two large-maximal
mixing patterns, $\theta_{13}(\Lambda_{\rm EW}) \sim 5^\circ$ can be 
radiatively produced in the MSSM. We have also demonstrated that it is 
possible to obtain $\theta_{13} \sim 5^\circ$ at $\Lambda_{\rm EW}$ from
$\theta_{13} = 0^\circ$ at $\Lambda_{\rm GUT}$ in the SM, if the
initial value of $\theta_{12}$ is large enough. As a useful by-product, 
the mass spectrum of three heavy Majorana neutrinos is determined and 
the cosmological baryon number asymmetry via leptogenesis is calculated.

Although the numerical examples presented in this paper are mainly for
the purpose of illustration, they are quite suggestive for model
building. Of course, only the future neutrino experiments can tell us 
how small $\theta_{13}$ is. But we believe that the radiative 
generation of $\theta_{13}$ from high to low energies is an interesting
theoretical approach towards understanding the smallness of  
$\theta_{13}$, and it might indicate some useful hints about the
underlying flavor symmetry which is associated with the dynamics of
lepton flavor mixing and CP violation. 
 
\vspace{0.5cm}

This work was supported in
part by the National Nature Science Foundation of China.

\newpage

\begin{table}
\caption{Numerical examples for radiative generation of
$\theta_{13}$ via the RGE evolution and seesaw threshold effects.
The cosmological baryon number asymmetry $Y_{\rm B}$ is also
computed. Note that we have taken 
$\theta_{12} =\theta_{23} =\delta =0^\circ$ for $V_\nu$ at 
$\Lambda_{\rm GUT}$.}
\vspace{0.3cm}
\begin{center}
\begin{tabular}{c|ccccc}
Inputs ($\Lambda_{\rm GUT}$)
& Pattern (A) & Pattern (B) & Pattern (C) & Pattern (D) & Model \\ \hline
$m_1$ (eV) & 0.12 & 0.08 & 0.06 & 0.06 & SM \\
           & 0.08 & 0.10 & 0.20 & 0.20 & MSSM \\ \hline
$\Delta m^2_{\rm sun}$ (${\rm eV}^2$)
 & 1.0$\times 10^{-4}$ & 1.8$\times 10^{-4}$ & 1.6$\times 10^{-4}$ 
& 1.2$\times 10^{-4}$ & SM \\
 & 2.1$\times 10^{-4}$ & 2.6$\times 10^{-4}$ & 5.8$\times 10^{-4}$ 
& 6.1$\times 10^{-4}$ & MSSM \\ \hline
$\Delta m^2_{\rm atm}$ (${\rm eV}^2$)
 & 7.9$\times 10^{-3}$ & 7.6$\times 10^{-3}$ & 8.0$\times 10^{-3}$ 
& 7.8$\times 10^{-3}$ & SM \\
 & 5.9$\times 10^{-3}$ & 5.7$\times 10^{-3}$ & 5.1$\times 10^{-3}$ 
& 5.1$\times 10^{-3}$ & MSSM \\ \hline
$\delta$ of $V_{\rm MNS}$
 & $70^\circ$ & $ 90^\circ$ & $ 75^\circ$ & $ 45^\circ$ & SM \\
 & $70^\circ$ & $100^\circ$ & $115^\circ$ & $115^\circ$ & MSSM \\
\hline
$\theta_{13}$ of $V_\nu$
 & $35^\circ$ & $50^\circ$ & $45^\circ$ & $45^\circ$ & SM \\
 & $27^\circ$ & $39^\circ$ & $45^\circ$ & $45^\circ$ & MSSM \\
\hline
$y_\nu$ & 0.9 & 0.8 & 0.9 & 0.9 & SM \\
        & 0.9 & 0.8 & 0.4 & 0.4 & MSSM \\ \hline
$r_1$ & 1/300 & 1/43 & 1/433 & 1/433 & SM \\
      & 1/300 & 1/43 & 1/433 & 1/433 & MSSM \\ \hline
$r_2$ & 1/19 & 1/15 & 1/19 & 1/19 & SM \\
      & 1/19 & 1/15 & 1/19 & 1/19 & MSSM \\ \hline
$m^{~}_H$ (GeV) & 120 & 120 & 120 & 120 & SM \\
$\tan\beta$ & 10 & 10 & 19 & 21 & MSSM \\ \hline
\hline
Outputs ($\Lambda_{\rm EW}$) & Pattern (A) & Pattern (B) & Pattern
(C) & Pattern (D) & Model \\ \hline
$m_1$ (eV) & 0.069 & 0.046 & 0.034 & 0.034 & SM \\
           & 0.053 & 0.067 & 0.14 & 0.14 & MSSM \\ \hline
$\Delta m^2_{\rm sun}$ (${\rm eV}^2$)
 & 6.9$\times 10^{-5}$ & 6.9$\times 10^{-5}$ & 6.9$\times 10^{-5}$ 
& 7.0$\times 10^{-5}$ & SM \\
 & 6.8$\times 10^{-5}$ & 7.3$\times 10^{-5}$ & 7.2$\times 10^{-5}$ 
& 6.9$\times 10^{-5}$ & MSSM \\ \hline
$\Delta m^2_{\rm atm}$ (${\rm eV}^2$)
 & 2.6$\times 10^{-3}$ & 2.6$\times 10^{-3}$ & 2.7$\times 10^{-3}$ 
& 2.6$\times 10^{-3}$ & SM \\
 & 2.6$\times 10^{-3}$ & 2.6$\times 10^{-3}$ & 2.6$\times 10^{-3}$ 
& 2.6$\times 10^{-3}$ & MSSM \\ \hline
$\theta_{13}$ of $V_{\rm MNS}$
 & $3.1^\circ$ & $1.6^\circ$ & $1.5^\circ$ & $1.4^\circ$ & SM \\
 & $2.2^\circ$ & $3.3^\circ$ & $4.7^\circ$ & $4.6^\circ$ & MSSM \\
\hline\hline
Outputs ($M_1$)
& Pattern (A) & Pattern (B) & Pattern (C) & Pattern (D) & Model \\ \hline
$M_1$ (GeV) & 3.0$\times 10^9$ & 1.8$\times 10^{11}$ & 2.9$\times 10^9$ 
& 2.2$\times 10^9$ & SM \\
            & 3.9$\times 10^9$ & 1.4$\times 10^{11}$ & 2.1$\times 10^8$ 
& 2.0$\times 10^8$ & MSSM \\ \hline
$M_2$ (GeV) & 8.4$\times 10^{11}$ & 9.9$\times 10^{11}$ 
& 1.0$\times 10^{12}$ & 8.9$\times 10^{11}$ & SM \\
            & 1.6$\times 10^{12}$ & 1.4$\times 10^{12}$ 
& 6.9$\times 10^{10}$ & 6.5$\times 10^{10}$ & MSSM \\ \hline
$M_3$ (GeV) & 7.9$\times 10^{13}$ & 1.2$\times 10^{14}$ 
& 1.7$\times 10^{14}$ & 2.7$\times 10^{14}$ & SM \\
            & 8.8$\times 10^{13}$ & 6.5$\times 10^{13}$ 
& 1.5$\times 10^{13}$ & 1.4$\times 10^{13}$ & MSSM \\ \hline
$Y_{\rm B}$ & 1.9$\times 10^{-12}$ & 8.9$\times 10^{-11}$ & 
1.7$\times 10^{-12}$ & 6.1$\times 10^{-13}$ & SM \\
            & 1.7$\times 10^{-12}$ & 7.5$\times 10^{-11}$ & 
1.6$\times 10^{-13}$ & 1.5$\times 10^{-13}$ & MSSM
\end{tabular}
\end{center}
\end{table}

\newpage

\begin{figure}
\begin{center}
\includegraphics[width=16cm,height=18cm]{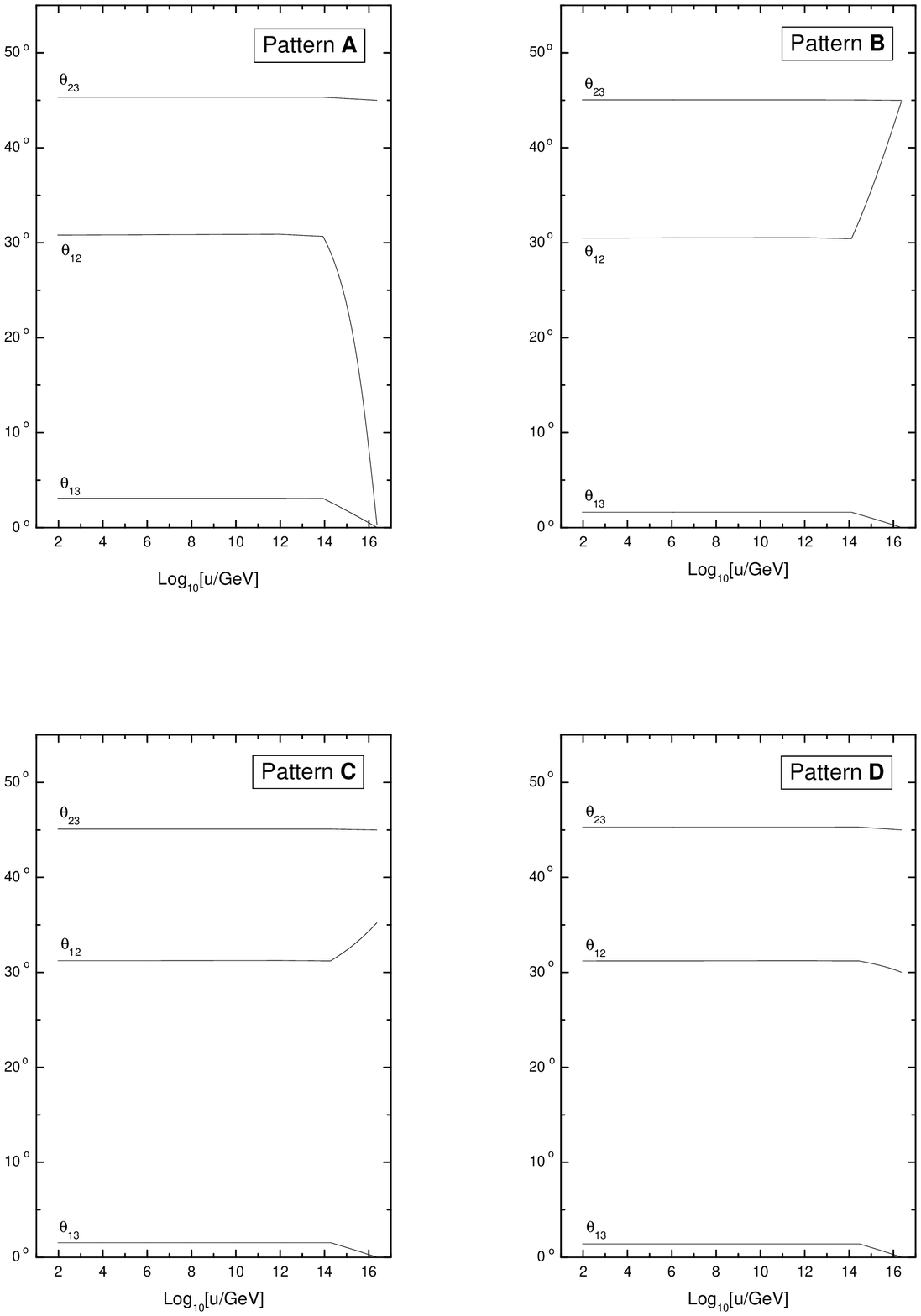}
\vspace{-0.4cm}
\caption{The RGE evolution of three lepton mixing
angles between $\Lambda_{\rm EW}$ and $\Lambda_{\rm GUT}$
in the SM, where the initial values of relevant parameters
are listed in Table 1.}
\end{center}
\end{figure}

\begin{figure}
\begin{center}
\includegraphics[width=16cm,height=18cm]{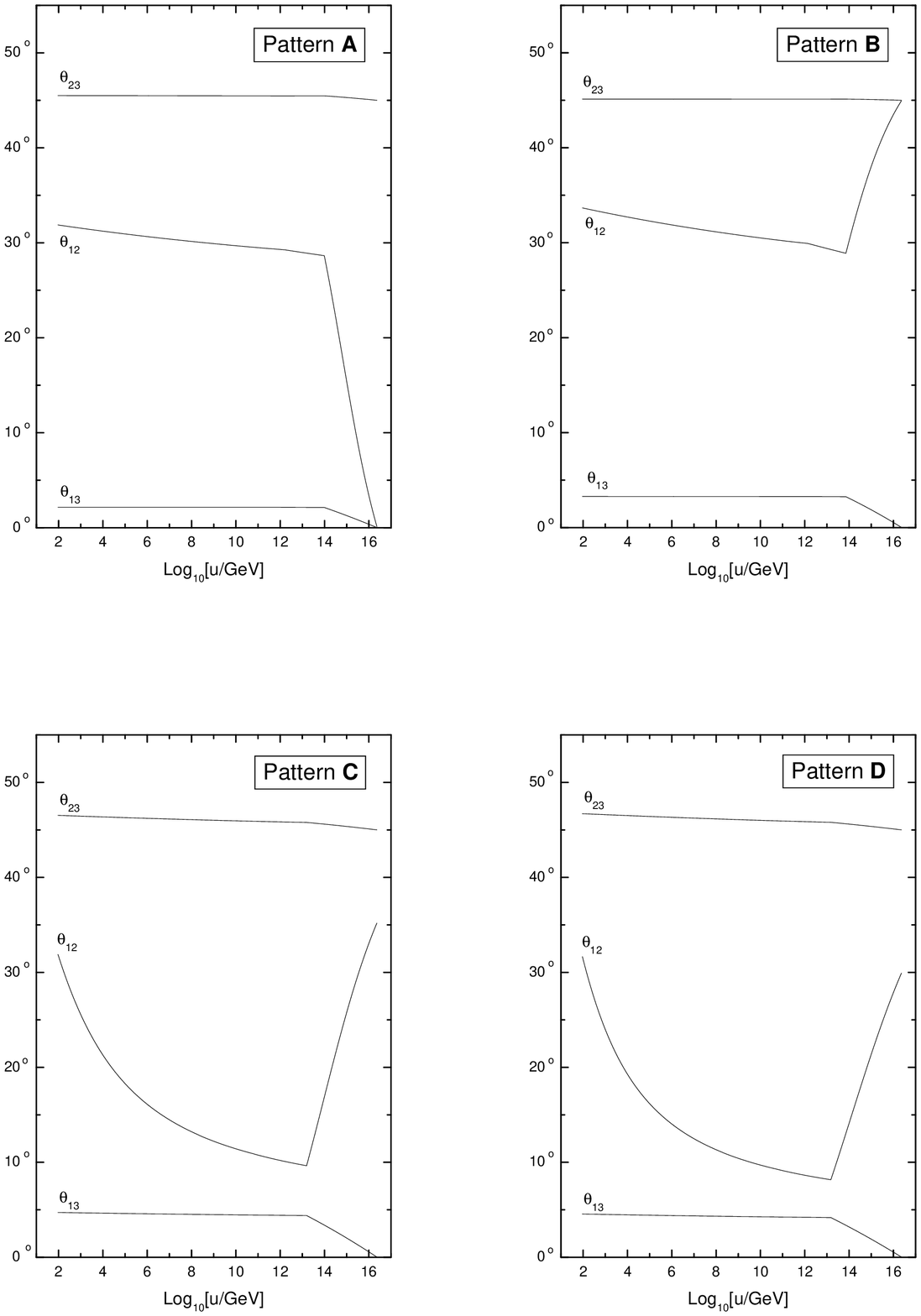}
\vspace{-0.4cm}
\caption{The RGE evolution of three lepton mixing
angles between $\Lambda_{\rm EW}$ and $\Lambda_{\rm GUT}$
in the MSSM, where the initial values of relevant parameters
are listed in Table 1.}
\end{center}
\end{figure}

\begin{figure}
\begin{center}
\includegraphics[width=12cm,height=18cm]{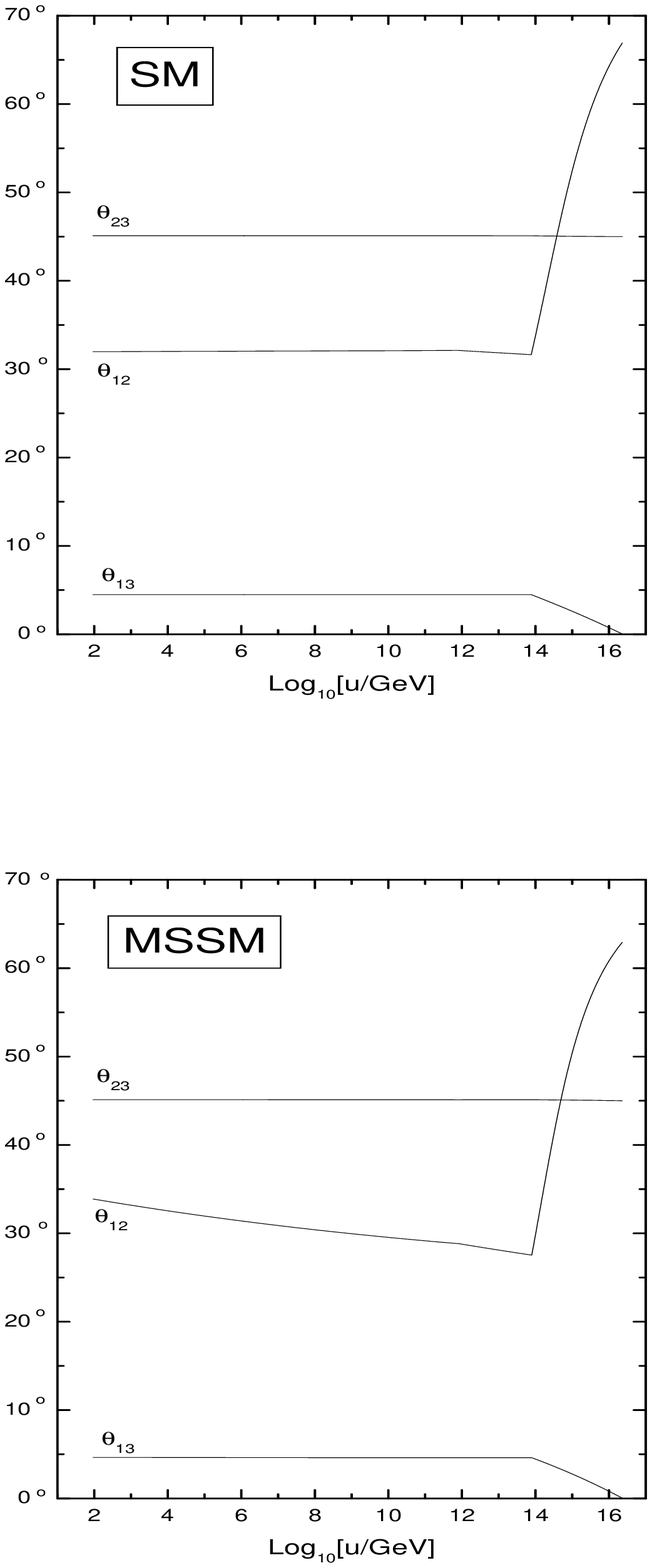}
\vspace{-0.4cm}
\caption{Running of three lepton mixing angles,
where $m_1 = 0.15$ eV, 
$\Delta m^2_{\rm sun} = 3.5 \times 10^{-4} ~{\rm eV}^2$,
$\Delta m^2_{\rm atm} = 7.6 \times 10^{-3} ~{\rm eV}^2$,
$y_\nu = 0.9$, $r_1 = 1/41$, $r_2 = 1/17$,
$\{\theta_{12}, \theta_{23}, \theta_{13}, \delta\}^{~}_{V_{\rm MNS}}
= \{67^\circ, 45^\circ, 0^\circ, 94^\circ\}$,
$\{\theta_{12}, \theta_{23}, \theta_{13}, \delta\}^{~}_{V_\nu}
= \{0^\circ, 0^\circ, 45^\circ, 0^\circ\}$ and 
$m^{~}_H = 120$ GeV have typically been input at
$\Lambda_{\rm GUT}$ in the SM; and 
$m_1 = 0.12$ eV, 
$\Delta m^2_{\rm sun} = 3.9 \times 10^{-4} ~{\rm eV}^2$,
$\Delta m^2_{\rm atm} = 5.4 \times 10^{-3} ~{\rm eV}^2$,
$y_\nu = 0.8$, $r_1 = 1/41$, $r_2 = 1/15$,
$\{\theta_{12}, \theta_{23}, \theta_{13}, \delta\}^{~}_{V_{\rm MNS}}
= \{63^\circ, 45^\circ, 0^\circ, 100^\circ\}$,
$\{\theta_{12}, \theta_{23}, \theta_{13}, \delta\}^{~}_{V_\nu}
= \{0^\circ, 0^\circ, 45^\circ, 0^\circ\}$ and 
$\tan\beta = 10$ have typically been input at
$\Lambda_{\rm GUT}$ in the MSSM.}
\end{center}
\end{figure}

\end{document}